# Global dynamics of neural mass models

G.K. Cooray[1,2,3], R.E. Rosch[2,4,5] and K.J. Friston[4]

[1] GOS-UCL Institute of Child Health, University College London, London, UK
[2] Great Ormond Street Hospital NHS Foundation Trust, London, UK
[3] Karolinska Institutet, Stockholm, Sweden
[4]The Wellcome Centre for Human Neuroimaging, Queen Square Institute of Neurology, University College London, London, UK
[5]MRC Centre for Neurodevelopmental Disorders, Institute of Psychiatry, Psychology and Neuroscience, King's College London, UK

Email: gerald.cooray@ki.se

**Abstract**

Neural mass models are used to simulate cortical dynamics and to explain the electrical and magnetic fields measured using electro- and magnetoencephalography. Simulations evince a complex phase-space structure for these kinds of models; including stationary points and limit cycles and the possibility for bifurcations and transitions among different modes of activity. This complexity allows neural mass models to describe the itinerant features of brain dynamics. However, expressive, nonlinear neural mass models are often difficult to fit to empirical data without additional simplifying assumptions: e.g., that the system can be modelled as linear perturbations around a fixed point. In this study we offer a mathematical analysis of neural mass models, specifically the canonical microcircuit model, providing analytical solutions describing dynamical itinerancy. We derive a perturbation analysis up to second order of the phase flow, together with adiabatic approximations. This allows us to describe amplitude modulations as gradient flows on a potential function of intrinsic connectivity. These results provide analytic proof-of-principle for the existence of semi-stable states of cortical dynamics at the scale of a cortical column. Crucially, this work allows for model inversion of neural mass models, not only around fixed points, but over regions of phase space that encompass transitions among semi or multi-stable states of oscillatory activity. In principle, this formulation of cortical dynamics may improve our understanding of the itinerancy that underwrites measures of cortical activity (through EEG or MEG). Crucially, these theoretical results speak to model inversion in the context of multiple semi-stable brain states, such as onset of seizure activity in epilepsy or beta bursts in Parkinson's disease.

# Introduction

The surface of the human brain is covered by a cortical layer of grey matter. This cortex contains histologically distinguishable layers with laminar-specific types of neural cells that show patterned interlaminar connectivity [1]. The cortex is also composed of cortical columns, that show denser interlaminar synaptic connectivity — within a column — than horizontally among adjacent columns [2]. Ensemble neuronal activity within each cortical column generates extracellular currents, which can be measured using microelectrodes with relatively high spatial precision; or more coarsely as net average currents through local field potential recordings [3].

There are numerous approaches to modelling the dynamics of neuronal activity at the scale of cortical columns. One-dimensional models describing simplified single-neuron dynamics with threshold firing; namely, *integrate-and-fire neurons* [4] can model the impact of interneuron connectivity on larger scale dynamics, such as those observed within and between cortical columns. However, these equations often result in integral-differential equations or partial differential equations, which are usually difficult to solve, at least in closed form, and especially when considering larger neuronal system [5,6].

Using mean-field approximations derived from statistical physics, the ensemble activity of cortical microcircuits comprising populations of distinct cell types have been modelled as neural mass point processes [7-12]. There are broadly two sets of neural mass models: (1) Conduction-based models based on the Hodgkin-Huxley model, with specific modelling of the nonlinear intrinsic dynamics of ion channels [13], which can allow for multi-compartment extensions [14,15]; and (2) convolution-based models where simple synaptic kernels are used to convolve presynaptic input to produce postsynaptic depolarization. These convolution based models are computationally efficient whilst allowing a range of complex dynamics to be simulated. The canonical microcircuit model (CMC) is a typical example of these convolution-based neural mass models. It comprises 4 populations and has been used extensively to model EEG and MEG data [16]. In this setting, an idealized cortical column is modelled using a set of $2^{nd}$-order differential equations with nonlinear coupling, using a sigmoid function to map the effect of the neuronal potential of one population on the depolarization of another.

In this study, we present an analysis of the dynamics of convolution-based neural mass models, specifically, the canonical microcircuit model [16]. When considering the spatially coarse-grained dynamics — typically recorded in human neurophysiological recordings — these kinds of models afford a balance between accuracy and complexity. In other words, they are sufficiently expressive to capture neuronal dynamics, at the scale of a cortical column, while being sufficiently simple to preclude overfitting, when used to explain empirical timeseries.

Electrical activity of the cortex can be readily recorded using electro- or magneto-encephalography (EEG/MEG). Visual analysis of the ensuing timeseries reveals a mixture of apparently stochastic features, such as paroxysmal discharges, intermixed with recurrent rhythms. The spectrum of rhythms and paroxysmal discharges vary in frequency, location and progression and are seen in a variety of states of the normal and diseased brain [3].

Quantitative analysis of brain states suggest that these recurrent, recognizable patterns can be modelled as semi-stable states, i.e., states with local but not global stability [10,17-20]. Importantly, these dynamical features are sensitive markers for whole-brain (dys)function and can indicate pharmacological and pathophysiological changes of neuronal connectivity at the synaptic level [21-23]. Thus, developing biophysical models of cortical dynamics may help identify the synaptic and connectivity processes that shape paroxysmal and time-varying dynamic features of brain activity.

The dynamical structure of these systems — such as neural mass models — can be characterized using a phase space representation. Simulation of the dynamics of neural mass models of the cortical column have disclosed complex structures in this phase space [24-26]. These include stationary points, limit cycles and chaotic attractors. Yet many popular current approaches of fitting neural mass models to empirical data, such as dynamic causal modelling, assume that the dynamics exist in a quasi steady state, without transitions between distinct dynamical modes [27]. There is therefore a current disconnect between the empirical observation of dynamically evolving brain states, and the steady-state assumption under which neural mass models are currently deployed to explain brain dynamics. Thus there is a need to develop the tools necessary to link biophysical models of cortical function to more complex, time-varying, and paroxysmal brain dynamics (e.g., seizure activity epilepsy or beta bursts in Parkinson's disease).

In what follows, we will use perturbative analysis to estimate closed form solutions for semi-stable (i.e., multi-stable) states of perturbations of the four-population canonical microcircuit (CMC) model. We use an adiabatic approximation to integrate over fast changing variables -- to derive phase-space dynamics that can be described by a potential field. To this relatively simple field equation we add Brownian noise, such that the dynamics of mean activity become gradient flows on the potential function. Crucially, this formulation can be used to generate key measures of itinerancy, such as the frequency and duration of transient oscillatory modes. Using this formulation, we briefly consider variational methods that could be used to estimate synaptic connectivity from empirical measures of itinerancy in the form of occupancy and mean exit times from distinct modes of activity. The theoretical results speak to a potential characterization of dynamical itinerancy seen in disorders of the brain—including epilepsy — in computationally tractable ways that can be incorporated in established (variational) model inversion schemes, such as dynamic causal modelling.

# Methods

This section describes the formal results upon which the proposed analysis rests. This analysis can be summarised, in narrative form, along the following lines:

- First, identify a sufficiently expressive neural mass model, whose equations of motion are parameterised in terms of synaptic connectivity.

- With a suitable transformation of variables, separate the fast (oscillatory) dynamics from slow (amplitude) fluctuations, in the spirit of an adiabatic approximation.

- Formulate the dynamics of the slow variables as a gradient flow on some potential function. This potential function can now be taken as the negative logarithm of a steady-state density; namely, the solution of the Fokker Planck equation describing density dynamics under random fluctuations.

- From the steady-state density, evaluate the probability of occupying various fixed points—in the state space of the slow variables—to generate the statistics of dwell times and transitions.

- Use the above steps as a generative model—that generates the statistics of dwell times from synaptic connectivity—to infer connectivity from empirically observed statistics of (slow) transitions among (fast) oscillatory modes of activity.

In what follows, we describe the mathematical basis of the steps above and present a brief, illustrative application, using simulated data to recover the connectivity giving rise to transient bursts of oscillatory activity.

## Equations of motion

The electrical activity of the cortical column can be modelled using the CMC-model as described in [16,27]. To keep the derivations simple, and to ensure readability, we will rename some parameters here, without changing the dynamics of the system. The model considers four distinct neuronal populations modelled using a set of four 2$^{nd}$ order differential equations. These ordinary differential equations can be written equivalently as eight first order equations.

$$\dot{p}_i = -\omega_i^2 q_i + \varepsilon \omega_i^2 \sum g_{ij} S(q_j) + \mu \omega_i^2 \sum g_{ij} P\left(\frac{p_j}{\omega_j}\right)$$

$$\dot{q}_i = p_i$$

Where the potential of the *i-th* neural population is given by $q_i$ and the rate of change of this is given by $p_i$ (i.e. a current term). We have introduced 2 variables $\mu$ and $\varepsilon$ to control the effect of perturbation of the non-connected model where the connectivity matrix is given by $g_{ij}$. A sigmoid function (*S*) is used to parametrize the interaction between the potential and current terms of the neuronal populations. The *P*-function parametrizes spike rate variability.

The original variables ($q_i$, $p_i$) can be written as a complex variable ($\in \mathbb{C}$) which will keep the derivations relatively concise and easy to follow.

$$z_i = q_i + i\frac{p_i}{\omega_i}$$

The closed loops of the unperturbed motion will be given by,

$$z_i z_i^* = R^2$$

The equation of unperturbed motion will be given by,

$$\dot{z}_i = -i\omega_i z_i$$

The solution will be given by,

$$z_i = R e^{-i\omega_i t}$$

The motion of the trajectories can then be given by the modulus (R) and argument ($\theta_i$) of the complex number ($z_i$), figure 1.

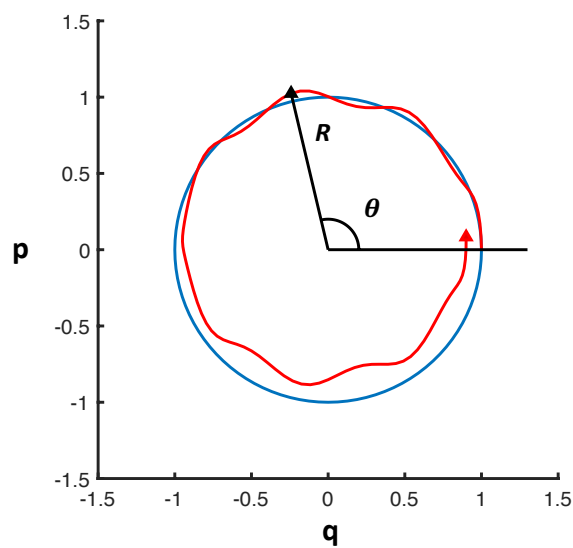

**Figure 1. Schematic figure showing the relation between the variables (q,p) and the amplitude and angle variables.** The blue curve shows the trajectory of an unperturbed trajectory (no damping and no coupling to other neuronal populations). The red curve depicts the trajectory of a perturbed path. The amplitude (*R*) and angle variable ($\theta$) for a point of the new trajectory (red) is shown.

We introduce a new phase variable $\varphi_i$ describing the deviation of the phase flow around its value when $\mu$ and $\varepsilon = 0$, i.e. no perturbation.

$$\theta_i = -\omega_i t + \varphi_i$$

The resulting equations of motion can then be given by,

$$\dot{z}_i = -i\omega_i z_i + i\varepsilon\omega_i \sum g_{ij} S\left(\frac{z_j + z_j^*}{2}\right) + i\mu\omega_i \sum g_{ij} P\left(\frac{z_j - z_j^*}{2i}\right) \qquad \text{Eq.1}$$

**Adiabatic approximation**

The above equations of motion can be simplified under the assumption that the time scale for phase dynamics (which is determined by $\omega_i$) is small in relation to the time scale for amplitude dynamics. This assumption has empirical validity when modelling cortical activity, as the amplitude modulation of cortical rhythms or spike amplitude evolves over a slower time scale (at least 10-100 times slower) than that of the phase dynamics [28,29].

In the above formulation this adiabatic assumption can be satisfied by limiting the range of $\varepsilon$ and $\mu$. The resulting adiabatic approximation is a simplified model of the amplitude variables (i.e., a 4-dimensional model), which permits further analysis of the dynamics in terms of closed form solutions. The usefulness of the ensuing adiabatic approximation can be assessed by the ability of the model to explain empirical data (see final section).

An adiabatic approximation can be specified with a time duration (*T*), where each neuronal population will complete—to 0$^{th}$ order (see *Perturbation analysis*)—a series of cycles. The angle variables move on a 4-dimensional torus, and *T* is chosen such that the motion over the torus completes (or approximately completes) one cycle and the angle variables can be eliminated from the dynamical flow [30].

$$T = least\ common\ denominator\ of\ \{\omega_i\}_{i=1,..,4}$$

**Perturbation analysis**

A perturbation analysis of Eq. 1 can now be pursued using the following expansion for $\varphi_i$ and $R_i$ (in modulus-argument form, or equivalently in polar coordinates).

$$z_i(t) = R_i(t,\varepsilon,\mu) e^{-i(\omega_i t + \varphi_i(t,\varepsilon,\mu))}$$

Where the two terms will be given by,

$$R_i(t,\varepsilon,\mu) = R_{i,0,0} + \sum \varepsilon^m \mu^n R_{i,m,n}(t)$$

$$\varphi_i(t,\varepsilon,\mu) = \sum \varepsilon^m \mu^n \varphi_{i,m,n}(t)$$

From the equation of motion we get the following,

$$\dot{z}_i(t) = \dot{R}_i e^{-i(\omega_i t + \varphi_i)} - iR_i(\omega_i + \dot{\varphi}_i)e^{-i(\omega_i t + \varphi_i)}$$

Which can be written as,

$$\left(\dot{R}_i(t,\varepsilon,\mu)e^{-i(\omega_i t+\varphi_i)} - iR_i(t,\varepsilon,\mu)\dot{\varphi}_i e^{-i(\omega_i t+\varphi_i)}\right)$$
$$= i\varepsilon\omega_i \sum g_{ij} S\left(\frac{z_j + z_j^*}{2}\right) + i\mu\omega_i^2 \sum g_{ij} P\left(\frac{z_j - z_j^*}{2i}\right)$$

The resulting equation can now be integrated over the closed (or almost closed) path over the torus, such that the angle variables are eliminated. To perform this integration we expand the perturbative functions S and P as polynomial series. We are now in a position to derive the change in $R_i$ to 1st order, by combining the results derived in the Appendix (A1-3):

$$\frac{dR_i}{dt} = \frac{\delta R_i}{T} = \frac{\mu R_{i,0,1}(T)}{T} = \mu\omega_i g_{ii} \sum_{r=1}^{\infty} B_r \frac{R_{i,0}^{2r-1}}{2^{2r-1}} \binom{2r-1}{r-1}$$

The dynamics of $R_i$ is determined by 2 coupling terms: the potential-to-current coupling (*S*) and the current-to-current coupling (*P*). The nonlinearity of the coupling functions (S and P) result in a series expansion in multiples of $R_i$. The complexity of the dynamics for the amplitude parameters is determined by the connectivity matrix and the coupling functions. These parameters are defined in the cortical column by reference to the synaptic connections between the neuronal populations, where the gain of each connection is given by $g_{ij}$ and their physiological effect on the target neuron parametrized in S and P. Note that the above derivation is valid for any set equal to or larger than two neuronal populations.

**Dynamics of amplitude modulations**

The electrical or current recordings of cortical activity is often modelled as a linear combination of the activity of the potential of each neuronal population [31]. This can be used to derive a 1-dimensional equation of motion for the amplitude dynamics of the recorded data. As can be seen above the amplitude dynamics will be governed by the self-connection terms, $g_{ii}$. The self-interactions within the cortical column will result in a set of non-interacting limit cycles (taking up-to 1st order contribution from the perturbation).

*Amplitude envelope dynamics*

The measured potential (e.g., EEG) is given by (assuming a linear lead field).

$$y(t) = \sum a_i R_i \cos\omega_i t \qquad \text{Eq.2}$$

The amplitude power can be estimated easily, provided the $\omega_i$ are distinct.

$$R^2 = \sum a_i{}^2 R_i{}^2$$

We can now normalize $a_i$ s.t

$$\sum_{i=1}^{n} a_i{}^2 = n$$

$R$ will then be the radius of a sphere if all $a_i$ are constant in the vector space given by $R_i$. The rate of change of the amplitude function $R$ is given as below.

$$R \frac{dR}{dt} = \sum a_i{}^2 R_i \frac{dR_i}{dt}$$

As we are interested in the average rate of change of $R$ we will take a spatial average over ellipsoids with constant $R^2$.

$$\frac{1}{A} \oiint R \frac{dR}{dt} = \frac{1}{A} \oiint \sum a_i{}^2 R_i \frac{dR_i}{dt}$$

$R$ will be constant on the surface ,

$$\frac{1}{A} \oiint \frac{dR}{dt} = \frac{1}{AR} \oiint \sum a_i{}^2 R_i \frac{dR_i}{dt}$$

Hyperspheroidal coordinates (in $\mathbb{R}^4$) are used for the integration (see Appendix A4). Let $R_s$ be the average of $R$ over the 3-sphere,

$$\frac{dR_s}{dt} = \frac{1}{AR} \oiint \sum a_i{}^2 R_i \frac{dR_i}{dt} = \mu \sum \omega_i g_{ii} \sum_{r=1}^{\infty} A_r \frac{4R^{2r-1}}{(2r+1)2^{2r} a_i{}^{2r}} \binom{2r-1}{r-1} \frac{(2r+1)!!}{(2r+2)!!} \qquad \text{Eq.3}$$

As was discussed in the appendix, cross connectivity terms (or off-diagonal terms) will affect the dynamics of $R_s$ when higher order terms are included in the perturbation expansion. The functional form of this equation means that the amplitude dynamics can be expressed as a gradient flow on a potential, *U* that depends upon the synaptic connections $g_{ij}$ and the coupling functions:

$$\partial_t I = -\nabla U$$

In other words, different connectivity matrixes and functions correspond to different potentials (*U*): see Figure 2. This spectrum then defines different possibilities of dynamics supported by the CMC model. Panel A shows minima at 0 and 2. Panel B has three stable points, at 0, 1 and 2 (see Appendix A1-3). This suggests that the CMC model—with certain synaptic connections—features continuous high amplitude oscillations and a point of stability at the origin. The oscillations would be a mixture of sinusoidal activity, with frequencies of the four neural populations, where this mixture is determined by the lead field (Eq. 2).

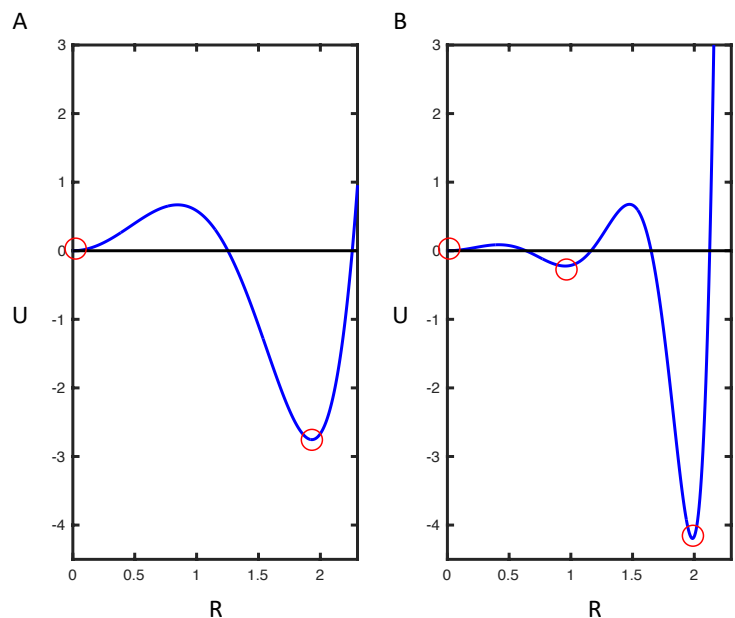

**Figure 2. Potential distributions for the dynamics of the total amplitude (*R*) for different connectivity between the neuronal subpopulations.** Here different connectivity matrices and couplings are selected to illustrate the spectrum of potential functions. Minima of $U$ in these functions are stable point attractors (A) The first example has two stable points at the origin and at R=2. (B) The second figure has three stable states at the origin and at R=1 and 2.

The potential distributions $U$ in Figure 2 are generated by one self-connectivity term in the superficial pyramidal cell layer. Up until this point, we have been considering deterministic dynamics in the absence of noise. In what follows, we show that the same dynamical structures persist in the presence of random fluctuations, via an analysis of the density dynamics.

A schematic of the steps taken to derive the dynamical equations of the amplitude modulation of the EEG is shown in figure 3.

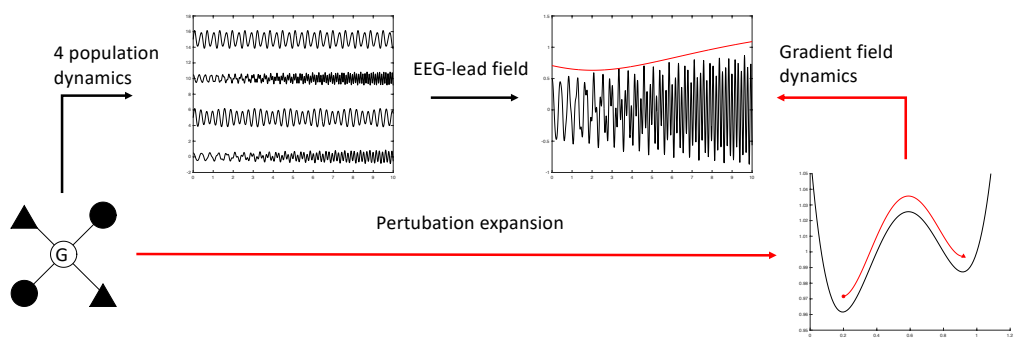

**Figure 3. A schematic figure of the steps taken to derive the amplitude dynamics.** The starting point is the 4 population CMC-model. The activity of the model generates activity for each of the four populations as shown after the arrow "4-population dynamics". The EEG lead field will then take a linear mixture of the activity of the 4 populations resulting in a "EEG" curve. The amplitude modulation of the activity is drawn by a red curve. The adiabatic and perturbation analysis discussed in this paper will define a gradient field as drawn after the arrow "perturbation expansion". The gradient field will depend on the connectivity matrix, G. A possible trajectory of the amplitude variable is shown as a red arrow moving from one stable state to another. The corresponding change in amplitude over time is shown after the arrow "gradient field dynamics" in red above the "EEG curve". The main contribution of this work is to find a direct relation between the intrinsic connectivity and the gradient field which in turn will define statistical measures such as the probability of being in a specific state or exit times from states.

**The Fokker Planck equation, stationary states and exit times**

Fast neuronal fluctuations can be easily modelled by adding innovations to the deterministic equations of motion [32,33]. For the potential flow of the amplitude dynamics derived above we can add a white (Wiener) noise process, resulting in a stochastic differential equation (SDE).

The trajectories are then described by the following SDE,

$$dI = -\nabla U dt + \sigma dB_t$$

In the above equation, *I* and $B_t$ are random variables while *U* and *σ* are real valued functions. The associated Fokker Planck equation, describing the density dynamics over the amplitude, is given by:

$$\partial_t \rho = \frac{\sigma^2}{2} \Delta \rho - \nabla.(\nabla U \rho)$$

The stationary state of this equation is given by the following, where *Z* is the partition function (or equivalently a normalizing constant for *p*).

$$p_{st} = \frac{1}{Z} exp\left\{-\frac{2U}{\sigma^2}\right\}$$

With a connectivity matrix that features several stable states, we can now estimate the probability of finding the dynamical system in that state by simply integrating the stationary probability density function (i.e., solution to the Fokker Planck equation above) over the interval defining the state: see Figure 4. This provides an analytic estimate of noise induced multistability (c.f., topological supersymmetry breaking), in terms of the probability of finding the amplitude dynamics near one of the fixed points or stable (oscillatory) states.

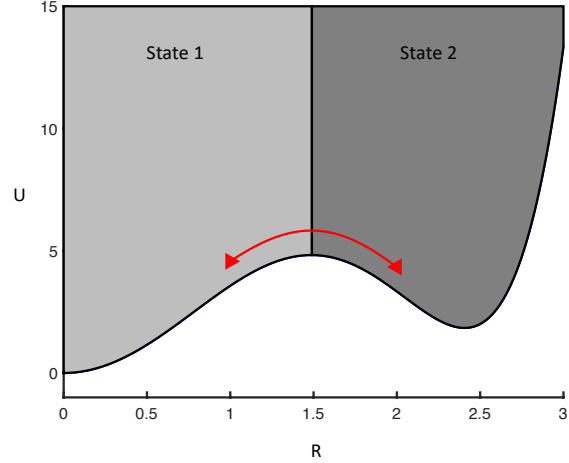

**Figure 4. Regions attracted to different minima of the potential** (shaded differently). The probability of finding the system in one of these stable states is given by integrating the probability density function over the attracting interval (state 1 = light grey), 0 to the first maxima ($R_{max}$) of the potential curve or $R_{max}$ to ∞ (state 2 = dark grey). The red arrow depicts how the trajectories pass between state 1 and 2 due to the underlying noise.

For the first state the probability is given by,

$$P\{state\ 1\} = \int_0^{R_{max}} \rho_{st} dI$$

The mean duration of time (*t*) the system is in a given state can be approximated using Kramer's escape rate,

$$\langle t \rangle = \frac{1}{2\pi\sqrt{-U''(x_{min})U''(x_{max})}} e^{\frac{2[U(x_{max})-U(x_{min})]}{\sigma^2}}$$

Finally, because we have closed form expressions for both the state probability and mean exit times, we are in a position to estimate the connectivity matrices from empirical measures of occupancy and mean exit times from discernible oscillatory states. In other words, the closed form expressions above provide the basis for a generative model that can be inverted given empirical measures of transitions among oscillatory modes evident in any empirical data. See Figure 5 for inversion of simulated data of exit times, for different noise amplitudes, using the variational procedures used in dynamic causal modelling. A simulated EEG trace is shown in Figure 6 illustrating how the system is moving between two semi-stable states resulting in a bursting appearance of the EEG.

In this proof of principle example, synthetic data were generated from using prior values of the connectivity in the CMC model provided by the SPM software. Inversion of the synthetic data was performed using standard (variational Laplace) inversion routines available in the SPM software package. This example shows that the connectivity parameters used to generate transitions among oscillatory states or modes can be recovered from mean exit times alone. The implication here is that it is possible to both explain and leverage multi-stability in terms of underlying synaptic connections and the amplitude of fast neuronal fluctuations.

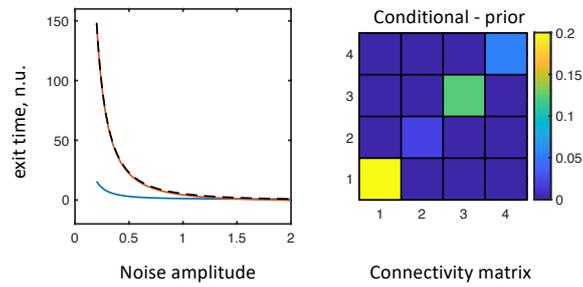

**Figure 5. Inversion of the generative model based on simulated data.** The left panel shows the mean exit time (on the vertical axis) from state 2 (Figure 4) for different noise amplitudes (plotted along the horizontal axis). The dotted black curve indicates the synthetic data that was inverted. The blue curve shows the expected data under the prior values of the connectivity matrix. The red curve indicates the expected exit times with the posterior values of the connectivity matrix. The right panel shows the change in the connectivity matrix between the prior and posterior estimates of the connectivity matrix, suggesting that the connectivity parameters used to generate the data can be recovered from mean exit time data features.

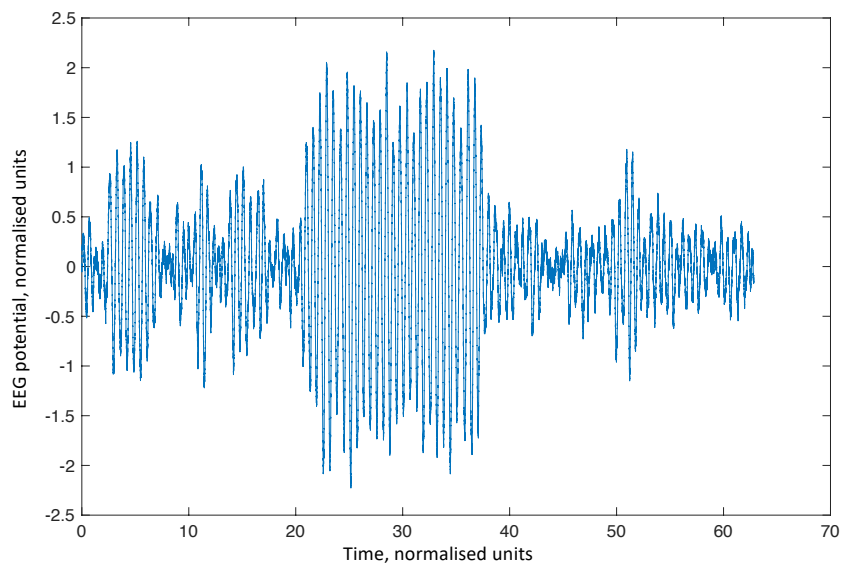

**Figure 6. Simulation of EEG trace with bi-stable activity.** The trace shows a simulated EEG trace with a connectivity matrix giving a bistable system as described in figure 4. The system is bursting with high amplitude oscillations (state 2) with shorter suppressions of attenuation (state 1) in between.

## Discussion

This paper has presented a formulation of multi-stable neuronal dynamics based on neural mass models—and an accompanying analysis framework—that enables inference about synaptic connectivity parameters that engender to observable transitions among oscillatory states. We have derived closed form expressions for the statistics of the semi-stable states, e.g. the state probability and the mean duration of a state.

Our work builds on both theoretical insights of prior neural mass modelling and empirical observations of transient brain states. Prior simulations of neural mass models have already shown the simultaneous presence of fixed point and limit cycle dynamics in the same cortical model [25,34]. Furthermore, empirically there are several instances—in both the healthy and diseased brain—where the mode of cortical and subcortical dynamics shifts repeatedly among states; including physiological rhythms like sleep spindles, mu-rhythms, and pathological states, such as epileptic seizure activity [3,35-39]. Indeed, the implicit itinerancy may be the hallmark of all self organising dynamical systems; ranging from the stochastic chaos that characterises nonequilibrium steady-state [40-47] through to heteroclinic cycles [48-50].

The neural mass models we have considered here show complex dynamics that can be divided into fast and slow variables. This separation in time scales allows for an adiabatic approximation, which eliminates the fast modes producing a new set of dynamical equations for the slow variables. This new set of slow variables now describes slowly evolving amplitude dynamics, rather than the fast oscillatory activity itself. This reduces the dimensionality of the system making it more tractable for simulation and inference. In this reduced system, several types of fixed points — representing stationary states or oscillatory limit cycles — can be characterised using measurable data features such as dwell and mean exit times.

In this work, we have presented closed form equations that link patterns of cortical microcircuit synaptic connectivity to statistical data features. This generative model allows for Bayesian model inversion and comparison, allowing empirical comparison of different mechanistic explanations of intermittent cortical dynamics. This might be particularly interesting when modelling relatively fast changing dynamics like on and off rhythmic spiking, subcortical beta rhythms, or seizure onset. As seizures progress there are changes in several variables including electrolytic or metabolic changes which would have a large but secondary effect on cortical dynamics and might change some key connectivity parameters, which would alter both fast and slow dynamics in the system described here [51]. These slower changes can also be modelled effectively using *Adiabatic DCM* where the effect of slow changes in synaptic parameters is included in the model inversion [52].

The amplitude dynamics of a single-source CMC model (i.e., cortical column) has been shown to be sufficiently approximated by a second order perturbation expansion. In contrast, extrinsic connections between cortical columns may require higher order perturbation terms and thus be of weaker strength compared to the intrinsic dynamics (at least within the parameter range where the perturbation expansion is valid). This is due to the number of neurons required to create an interconnected loop between two cortical columns [2,53]. The intrinsic (i.e., within-source or column) dynamics is thus instantiated in cortical columns through the intrinsic connectivity matrix, while the extrinsic connectivity results in a weak but richer set of dynamics, as the extrinsic (i.e., between-source or column) connectivity can be diverse. This is in line with the architectural setup of cortical processing, where cortical columns represent the smallest units that process data as conditionally independent units and interact with other columns, using a weaker set of interactions [54]. We will further develop the perturbation theory presented here to include higher order terms to see whether the interactions between cortical columns can be characterised.

Our perturbation expansion is sufficient to explain semi-stable states of the cortical modes of activity (i.e. stationary states and limit cycles of the full dynamics) within cortical columns in the absence of changes of connectivity. Using a perturbation expansion, we have managed to derive the characteristics of semi-stable states of the cortical columns from the connectivity matrix describing connectivity within the cortical column. This is an important step, as we can now map statistical characterisations of complex cortical dynamics onto synaptic connectivity. Understanding larger cortical networks from a modelling perspective can be further implemented using a discrete simplification of the above model. It is possible — through a projection mapping of the dynamics of the cortical columns onto the semi-stable state it occupies at any given time — to simplify things considerably. This should preserve the important link between connectivity and the semi-stable states that are generated. Connecting different cortical columns could then result in a lattice model (similar to an Ising or Pott's model) of the cortex, where it would be relatively straightforward to link the intrinsic and extrinsic connectivity to measurable dynamics [55,56]. Such an expansion of our work would prove particularly relevant to the characterisation of epileptogenic networks and their role in seizure initiation and spread. Furthermore, the effect of medication could be addressed in terms of the changes in the intrinsic connectivity and in ceiling changes in network dynamics. Finally, the effect of epilepsy surgery or disconnection surgery could be modelled by changes in extrinsic connectivity; c.f. [57-59].

The approach illustrated here may have particular relevance for our understanding of paroxysmal disorders of cortical function such as epilepsy and epileptic seizures. The epilepsies are characterised by recurrent, sudden onset of rhythmic epileptic activity, on a background of spontaneous 'normal' interictal ongoing activity. In certain cases, cortical activity in epilepsy can alternate rapidly between more than two states, 1) rhythmic spiking, 2) seizure activity and 3) spontaneous background activity [17,60]. The transition between these states has been described in terms of phase transitions and bifurcations, where the synaptic parameters that shape microcircuit dynamics exceed a threshold, resulting in a sudden change in phase space dynamics [29]. Our model, similar to previous modelling endeavours, offers an alternative perspective in that it does not require a critical transition (as in transcritical bifurcations) to explain the change in cortical activity [61]. Rather, the mechanism resulting in intermittently observable dynamic phenomena is predicated on noise induced itinerancy, of the kind found in stochastic chaos and topological supersymmetry breaking. In other words, the stochastic nature of the model will move the dynamics of the system between semi-stable states.

Observations of intermittent dynamic features could—in principle—give support for either bifurcations or transitions between semi-stable states. Note that we do not provide any evidence for the absence of bifurcations in these intermittent systems, as these still exist within the framework presented (i.e. changing the connectivity matrix can still cause transcritical bifurcations): rather that these are not always required to explain observable state transitions. To distinguish between the two proposals; bifurcations or semi-stable state transitions, the statistics of transitions among states may need to be further explored. It is through tools such as the invertible, reduced models presented here that future work will be able to specifically address these questions empirically; i.e., through the (Bayesian) model comparison of models with and without changes in connectivity.

In summary, we have introduced the theoretical backdrop for a model of cortical multi-stability and itinerancy that could be used to infer the underlying synaptic connectivity from statistical descriptions of transient dynamic features. We hope to elaborate the theoretical work presented above for the characterisation of global dynamics based on measurable electromagnetic responses in health and disease.

**Funding statement**

KJF is supported by funding for the Wellcome Centre for Human Neuroimaging (Ref: 205103/Z/16/Z) and a Canada-UK Artificial Intelligence Initiative (Ref: ES/T01279X/1). RER is supported by funding from the Wellcome Trust (209164/Z/17/Z). KJF and RER are supported by the European Union's Horizon 2020 Framework Programme for Research and Innovation under the Specific Grant Agreement No. 945539 (Human Brain Project SGA3).

# References

[1] Jiang X, Shen S, Cadwell CR, Berens P, Sinz F, Ecker AS, Patel S, Tolias AS. Principles of connectivity among morphologically defined cell types in adult neocortex. Science. 2015 Nov 27;350(6264):aac9462.

[2] Douglas R, Martin K, Whitteridge D, A Canonical Microcircuit for Neocortex, Neural Computation, 1989

[3] Niedermeyer E, da Silva FL, editors. Electroencephalography: basic principles, clinical applications, and related fields. Lippincott Williams & Wilkins; 2005.

[4] Abbott LF, Kepler TB. Model neurons: from Hodgkin Huxley to Hopfield. In Statistical mechanics of neural networks 1990 (pp. 5-18). Springer, Berlin, Heidelberg.

[5] Coombes S. Waves, bumps, and patterns in neural field theories. Biological cybernetics. 2005 Aug;93(2):91-108.

[6] Brunel, N., and Hakim, V. (1999). Fast global oscillations in networks of integrate-and-fire neurons with low firing rates. *Neural Comput.* 11, 1621–1671.

[7] Jansen BH, Rit VG. Electroencephalogram and visual evoked potential generation in a mathematical model of coupled cortical columns. Biological cybernetics. 1995 Sep;73(4):357-66.

[8] David O, Friston KJ. A neural mass model for MEG/EEG: coupling and neuronal dynamics. NeuroImage. 2003 Nov 1;20(3):1743-55.

[9] Wilson HR, Cowan JD. A mathematical theory of the functional dynamics of cortical and thalamic nervous tissue. Kybernetik. 1973 Sep;13(2):55-80.

[10] Lopes da Silva FH, Hoeks A, Smits H, and Zetterberg LH. Model of brain rhythmic activity: The alpha-rhythm of the thalamus. *Kybernetik*, 15:27–37, 1974.

[11] Wendling F, Benquet P, Bartolomei F, and Jirsa V. Computational models of epilepti- form activity. *Journal of Neuroscience Methods*, 260:233–251, 2016.

[12] Dafilis MP, Frascoli F, Cadusch PJ, and Liley DTJ. Chaos and generalised multista- bility in a mesoscopic model of the electroencephalogram. *Physica D*, 238:1056–1060, 2009.

[13] Hodgkin AL, Huxley AF. A quantitative description of membrane current and its application to conduction and excitation in nerve. The Journal of physiology. 1952 Aug 28;117(4):500.

[14] Rall W. Branching dendritic trees and motoneuron membrane resistivity. Experimental neurology. 1959 Nov 1;1(5):491-527.

[15] Traub RD. Motorneurons of different geometry and the size principle. Biological Cybernetics. 1977 Sep;25(3):163-76.

[16] Bastos AM, Usrey WM, Adams RA, Mangun GR, Fries P, Friston KJ. Canonical microcircuits for predictive coding. Neuron. 2012 Nov 21;76(4):695-711.

[17] Cooray CN, Carvalho A, Cooray GK. Noise induced quiescence of epileptic spike generation in patients with epilepsy. Journal of Computational Neuroscience. 2021 Feb;49(1):57-67.

[18] Wendling F, Bellanger JJ, Bartolomei F, Chauvel P. Relevance of nonlinear lumped-parameter models in the analysis of depth-EEG epileptic signals. Biological cybernetics. 2000 Sep;83(4):367-78.

[19]Freeman WJ. Models of the dynamics of neural populations. Electroencephalography and clinical neurophysiology. Supplement. 1978 Jan 1(34):9-18.

[20] Liu F, Wang J, Liu C, Li H, Deng B, Fietkiewicz C, Loparo KA. A neural mass model of basal ganglia nuclei simulates pathological beta rhythm in Parkinson's disease. Chaos: An Interdisciplinary Journal of Nonlinear Science. 2016 Dec 15;26(12):123113.

[21] Boly M, Moran R, Murphy M, Boveroux P, Bruno MA, Noirhomme Q, Ledoux D, Bonhomme V, Brichant JF, Tononi G, Laureys S. Connectivity changes underlying spectral EEG changes during propofol-induced loss of consciousness. Journal of Neuroscience. 2012 May 16;32(20):7082-90.

[22] Smailovic U, Koenig T, Kåreholt I, Andersson T, Kramberger MG, Winblad B, Jelic V. Quantitative EEG power and synchronization correlate with Alzheimer's disease CSF biomarkers. Neurobiology of aging. 2018 Mar 1;63:88-95.

[23] Willoughby JO, Fitzgibbon SP, Pope KJ, Mackenzie L, Medvedev AV, Clark CR, Davey MP, Wilcox RA. Persistent abnormality detected in the non-ictal electroencephalogram in primary generalised epilepsy. Journal of Neurology, Neurosurgery & Psychiatry. 2003 Jan 1;74(1):51-5.

[24] Ermentrout GB, Kopell N. Parabolic bursting in an excitable system coupled with a slow oscillation. SIAM journal on applied mathematics. 1986 Apr;46(2):233-53.

[25] Grimbert F, Faugeras O. Bifurcation analysis of Jansen's neural mass model. Neural computation. 2006 Dec;18(12):3052-68.

[26] Touboul J, Wendling F, Chauvel P, Faugeras O. Neural mass activity, bifurcations, and epilepsy. Neural computation. 2011 Dec 1;23(12):3232-86.

[27] Moran RJ, Pinotsis DA, Friston KJ. Neural masses and fields in dynamic causal modeling. Frontiers in computational neuroscience. 2013 May 28;7:57.

[28] Novak P, Lepicovska V, Dostalek C. Periodic amplitude modulation of EEG. Neuroscience letters. 1992 Mar 2;136(2):213-5.

[29] Jirsa VK, Stacey WC, Quilichini PP, Ivanov AI, Bernard C. On the nature of seizure dynamics. Brain. 2014 Aug 1;137(8):2210-30.

[30] Arnold VI. Mathematical methods of classical mechanics. Springer Science & Business Media; 2013 Apr 9.

[31] Mosher JC, Leahy RM, Lewis PS. EEG and MEG: forward solutions for inverse methods. IEEE Transactions on biomedical engineering. 1999 Mar;46(3):245-59.

[32] Friston KJ, Li B, Daunizeau J, Stephan KE. Network discovery with DCM. Neuroimage. 2011 Jun 1;56(3):1202-21.

[33] Oksendal B. Stochastic differential equations: an introduction with applications. Springer Science & Business Media; 2013 Mar 9.

[34] Spiegler A, Kiebel SJ, Atay FM, Knösche TR. Bifurcation analysis of neural mass models: Impact of extrinsic inputs and dendritic time constants. NeuroImage. 2010 Sep 1;52(3):1041-58.

[35] Alvarado-Rojas C, Valderrama M, Fouad-Ahmed A, Feldwisch-Drentrup H, Ihle M, Teixeira CA, Sales F, Schulze-Bonhage A, Adam C, Dourado A, Charpier S. Slow modulations of high-frequency activity (40–140 Hz) discriminate preictal changes in human focal epilepsy. Scientific reports. 2014 Apr 1;4(1):1-9.

[36] Brown P. Bad oscillations in Parkinson's disease. Parkinson's Disease and Related Disorders. 2006:27-30.

[37] Canolty RT, Edwards E, Dalal SS, Soltani M, Nagarajan SS, Kirsch HE, Berger MS, Barbaro NM, Knight RT. High gamma power is phase-locked to theta oscillations in human neocortex. science. 2006 Sep 15;313(5793):1626-8.

[38] Cooray GK, Sengupta B, Douglas P, Englund M, Wickstrom R, Friston K. Characterising seizures in anti-NMDA-receptor encephalitis with dynamic causal modelling. Neuroimage. 2015 Sep 1;118:508-19.

[39] Sterio D, Berić A, Dogali M, Fazzini E, Alfaro G, Devinsky O. Neurophysiological properties of pallidal neurons in Parkinson's disease. Annals of Neurology: Official Journal of the American Neurological Association and the Child Neurology Society. 1994 May;35(5):586-91.

[40] Ao P. Potential in stochastic differential equations: novel construction. Journal of physics A: mathematical and general. 2004 Jan 7;37(3):L25.

[41] Agarwal S, Wettlaufer JS. Maximal stochastic transport in the Lorenz equations. Physics Letters A. 2016 Jan 8;380(1-2):142-6.

[42] Friston K, Heins C, Ueltzhöffer K, Da Costa L, Parr T. Stochastic chaos and markov blankets. Entropy. 2021 Sep 17;23(9):1220.

[43] Jiang, Da-Quan, and Donghua Jiang. *Mathematical theory of nonequilibrium steady states: on the frontier of probability and dynamical systems*. Springer Science & Business Media, 2004.

[44] Kim EJ. Investigating information geometry in classical and quantum systems through information length. Entropy. 2018 Aug 3;20(8):574.

[45] Lasota A, Mackey MC. Chaos, fractals, and noise: stochastic aspects of dynamics. Springer Science & Business Media; 1998 Apr 1.

[46] Nicolis G, Prigogine I. Self-organization in nonequilibrium systems: from dissipative structures to order through fluctuations.

[47] Yan H, Zhao L, Hu L, Wang X, Wang E, Wang J. Nonequilibrium landscape theory of neural networks. Proceedings of the National Academy of Sciences. 2013 Nov 5;110(45):E4185-94.

[48] Friston KJ. The labile brain. II. Transients, complexity and selection. Philosophical Transactions of the Royal Society of London. Series B: Biological Sciences. 2000 Feb 29;355(1394):237-52.

[49] Rabinovich M, Huerta R, Laurent G. Transient dynamics for neural processing. Science. 2008 Jul 4;321(5885):48-50.

[50] Rabinovich MI, Afraimovich VS, Bick C, Varona P. Information flow dynamics in the brain. Physics of life reviews. 2012 Mar 1;9(1):51-73.

[51] Meyer JS, Gotoh F, Favale E. Cerebral metabolism during epileptic seizures in man. Electroencephalography and Clinical Neurophysiology. 1966 Jul 1;21(1):10-22.

[52] Jafarian A, Zeidman P, Wykes RC, Walker M, Friston KJ. Adiabatic dynamic causal modelling. NeuroImage. 2021 Sep 1;238:118243.

[53] Kiebel SJ, Garrido MI, Moran RJ, Friston KJ. Dynamic causal modelling for EEG and MEG. Cognitive neurodynamics. 2008 Jun;2(2):121-36.

[54] Creutzfeldt OD. Generality of the functional structure of the neocortex. Naturwissenschaften. 1977 Oct;64(10):507-17.

[55] Ising E. *Beitrag zur theorie des ferro-und paramagnetismus* (Doctoral dissertation, Grefe & Tiedemann).

[56] Potts RB. Some generalized order-disorder transformations. InMathematical proceedings of the cambridge philosophical society 1952 Jan (Vol. 48, No. 1, pp. 106-109). Cambridge University Press.

[57] Jirsa VK, Friedrich R, Haken H, Kelso JS. A theoretical model of phase transitions in the human brain. Biological cybernetics. 1994 May;71(1):27-35.

[58] Proix T, Jirsa VK, Bartolomei F, Guye M, Truccolo W. Predicting the spatiotemporal diversity of seizure propagation and termination in human focal epilepsy. Nature communications. 2018 Mar 14;9(1):1-5.

[59] Roy D, Sigala R, Breakspear M, McIntosh AR, Jirsa VK, Deco G, Ritter P. Using the virtual brain to reveal the role of oscillations and plasticity in shaping brain's dynamical landscape. Brain connectivity. 2014 Dec 1;4(10):791-811.

[60] Lopes da Silva FH, Blanes W, Kalitzin SN, Parra J, Suffczynski P, Velis DN. Dynamical diseases of brain systems: different routes to epileptic seizures. IEEE transactions on biomedical engineering. 2003 May 13;50(5):540-8.

[61] Goodfellow M, Schindler K, Baier G. Intermittent spike–wave dynamics in a heterogeneous, spatially extended neural mass model. NeuroImage. 2011 Apr 1;55(3):920-32.

## Appendix

The complex function for the trajectory will be given by (in modulus-argument form, or equivalently in polar coordinates),

$$z_i(t) = R_i(t, \varepsilon, \mu) e^{-i(\omega_i t + \varphi_i((t, \varepsilon, \mu)))}$$

The perturbation expansion will be given by,

$$R_i(t, \varepsilon, \mu) = R_{i,0,0} + \sum \varepsilon^m \mu^n R_{i,m,n}(t)$$

$$\varphi_i(t, \varepsilon, \mu) = \sum \varepsilon^m \mu^n \varphi_{i,m,n}(t)$$

The equation of motion will be given by,

$$\left( \dot{R}_i(t, \varepsilon, \mu) e^{-i(\omega_i t + \varphi_i)} - i R_i(t, \varepsilon, \mu) \dot{\varphi}_i e^{-i(\omega_i t + \varphi_i)} \right) \qquad \text{Eq. A1}$$
$$= i\varepsilon \omega_i \sum g_{ij} S\left( \frac{z_j + z_j^*}{2} \right) + i\mu \omega_i \sum g_{ij} P\left( \frac{z_j - z_j^*}{2i} \right)$$

The sigmoid function (tanh) can be expanded as,

$$S\left( \frac{z_j + z_j^*}{2} \right) = \sum_{r=1}^{\infty} A_r \left( \frac{z_j + z_j^*}{2} \right)^{2r-1}$$

$$= \sum_{r=1}^{\infty} A_r \frac{\left[ R_j(t, \varepsilon, \mu) e^{-i(\omega_j t + \varphi_j)} + R_j(t, \varepsilon, \mu) e^{i(\omega_j t + \varphi_j)} \right]^{2r-1}}{2^{2r-1}}$$

$$= \sum_{r=1}^{\infty} A_r \frac{\left( R_{j,0,0} + \sum \varepsilon^n \mu^m R_{j,n,m}(t) \right)^{2r-1}}{2^{2r-1}} e^{-i(2r-1)(\omega_j t + \varphi_j)}$$

$$* \left[ 1 + e^{i2(\omega_j t + \varphi_j)} \right]^{2r-1}$$

$$= \sum_{r=1}^{\infty} A_r \frac{\left( R_{j,0,0} + \sum \varepsilon^n \mu^m R_{j,n,m}(t) \right)^{2r-1}}{2^{2r-1}} e^{-i(2r-1)(\omega_j t + \varphi_j)}$$

$$* \left[ \sum_{l=0}^{2r-1} \binom{2r-1}{l} e^{i2l(\omega_j t + \varphi_j)} \right]$$

$$= \sum_{r=1}^{\infty} A_r \frac{\left(R_{j,0,0} + \sum \varepsilon^n \mu^m R_{j,n,m}(t)\right)^{2r-1}}{2^{2r-1}}$$

$$* \left[\sum_{l=0}^{2r-1} \binom{2r-1}{l} e^{-i((2r-1)-2l)(\omega_j t)} \left[\sum_{k=0}^{\infty} \frac{(-i(2r-1-2l)\varphi_j)^k}{k!}\right]\right]$$

The *P* function (current-to-current coupling) can be expanded as,

$$P\left(\frac{z_j - z_j^*}{2i}\right) = \sum_{r=1}^{\infty} B_r \left(\frac{z_j - z_j^*}{2i}\right)^{2r-1}$$

$$= i \sum_{r=1}^{\infty} B_r \frac{\left[R_j(t,\mu) e^{-i(\omega_j t + \varphi_j)} - R_j(t,\mu) e^{i(\omega_j t + \varphi_j)}\right]^{2r-1}}{2^{2r-1}(-1)^r}$$

$$= i \sum_{r=1}^{\infty} B_r \frac{\left(R_{j,0} + \sum \mu^n R_{j,n}(t)\right)^{2r-1}}{2^{2r-1}(-1)^r} e^{-i(2r-1)(\omega_j t + \varphi_j)} * \left[1 - e^{i2(\omega_j t + \varphi_j)}\right]^{2r-1}$$

$$= i \sum_{r=1}^{\infty} B_r \frac{\left(R_{j,0} + \sum \mu^n R_{j,n}(t)\right)^{2r-1}}{2^{2r-1}(-1)^r} e^{-i(2r-1)(\omega_j t + \varphi_j)}$$

$$* \left[\sum_{l=0}^{2r-1} (-1)^l \binom{2r-1}{l} e^{i2l(\omega_j t + \varphi_j)}\right]$$

$$= i \sum_{r=1}^{\infty} (-1)^r B_r \frac{\left(R_{j,0} + \sum \mu^n R_{j,n}(t)\right)^{2r-1}}{2^{2r-1}}$$

$$* \left[\sum_{l=0}^{2r-1} (-1)^l \binom{2r-1}{l} e^{-i((2r-1)-2l)(\omega_j t)} \left[\sum_{k=0}^{\infty} \frac{(-i(2r-1-2l)\varphi_j)^k}{k!}\right]\right]$$

Expanding the LHS of Eq A1.

$$\left(\dot{R}_i(t,\varepsilon,\mu)e^{-i(\omega_i t+\varphi_i)} - iR_i(t,\varepsilon,\mu)\dot{\varphi}_i e^{-i(\omega_i t+\varphi_i)}\right) \qquad \text{Eq. A2}$$

$$= \left[\sum_{k=0}^{\infty}\frac{(-i\varphi_i)^k}{k!}\right]\left(\sum \varepsilon^n \mu^m \dot{R}_{i,n,m}(t)\right)e^{-i\omega_i t} - i\left(R_{i,0,0} + \sum \varepsilon^n \mu^m R_{i,n,m}(t)\right)$$

$$* \left(\sum \varepsilon^n \mu^m \dot{\varphi}_{i,n,m}(t)\right)\left[\sum_{k=0}^{\infty}\frac{(-i\varphi_i)^k}{k!}\right]e^{-i\omega_i t}$$

The RHS of Eq. A1, after multiplying with $e^{i\omega_i t}$, will be given by,

$$i\omega_i \varepsilon e^{i\omega_i t}\sum g_{ij}\sum_{r=1}^{\infty}A_r\frac{\left(R_{j,0,0}+\sum\varepsilon^n\mu^m R_{j,n,m}(t)\right)^{2r-1}}{2^{2r-1}} \qquad \text{Eq. A3}$$

$$* \left[\sum_{l=0}^{2r-1}\binom{2r-1}{l}e^{-i((2r-1)-2l)(\omega_j t)}\left[\sum_{k=0}^{\infty}\frac{(-i(2r-1-2l)\varphi_j)^k}{k!}\right]\right]$$

$$+ i\omega_i\mu e^{i\omega_i t}\sum g_{ij}\sum_{r=1}^{\infty}(-1)^r B_r\frac{\left(R_{j,0,0}+\sum\mu^n R_{j,0,n}(t)\right)^{2r-1}}{2^{2r-1}}$$

$$* \left[\sum_{l=0}^{2r-1}(-1)^l\binom{2r-1}{l}e^{-i((2r-1)-2l)(\omega_j t)}\left[\sum_{k=0}^{\infty}\frac{(-i(2r-1-2l)\varphi_j)^k}{k!}\right]\right]$$

**A1. Potential-to-current coupling**
We will investigate the effect of potential-to-current coupling by keeping $\mu = 0$, keeping only the *S*-function.

*A1.1 Expanding the $\varepsilon^1$ term*

Keeping only $\varepsilon^1$ terms for the LHS (Eq. A2) will give,

$$\varepsilon^1\dot{R}_{i,1,0}(t) - iR_{i,0,0}\varepsilon^1\dot{\varphi}_{i,1,0}(t)$$

Keeping only $\varepsilon^1$ terms for the RHS (Eq. A3) will give,

$$i\varepsilon\omega_i\sum g_{ij}\sum_{r=1}^{\infty}A_r\frac{R_{j,0,0}^{2r-1}}{2^{2r-1}}\left[\sum_{l=0}^{2r-1}\binom{2r-1}{l}e^{-i((2r-1)-2l)(\omega_j t)+i\omega_i t}\right]$$

Equating these terms will give,

$$\dot{R}_{i,1,0}(t) - iR_{i,0,0}\dot{\varphi}_{i,1,0}(t)$$
$$= i\omega_i \sum g_{ij} \sum_{r=1}^{\infty} A_r \frac{R_{j,0,0}^{2r-1}}{2^{2r-1}} e^{-i(2r-1)(\omega_j t)+i\omega_i t} \left[ \sum_{l=0}^{2r-1} \binom{2r-1}{l} e^{i2l(\omega_j t)} \right]$$
$$\equiv W_{i,1,0}$$

We can then calculate the following,

$$\dot{R}_{i,1,0}(t) = \frac{W_{i,1,0} + W_{i,1,0}^*}{2}$$

$$\dot{\varphi}_{i,1,0}(t) = i\frac{W_{i,1,0} - W_{i,1,0}^*}{2R_{i,0,0}}$$

Integrating the terms will give,

$$R_{i,1,0}(t) = \int_0^t \dot{R}_{i,1,0}(t') = \int_0^t \frac{W_{i,1,0} + W_{i,1,0}^*}{2} \qquad \text{Eq. A4}$$

$$\varphi_{i,1,0}(t) = \int_0^t \dot{\varphi}_{i,1,0}(t') = i\int_0^t \frac{W_{i,1,0} - W_{i,1,0}^*}{2R_{i,0,0}} \qquad \text{Eq. A5}$$

The integral on *W* will be given by,

$$\int_0^t W_{i,1,0} = i\omega_i \sum g_{ij} \sum_{r=1}^{\infty} A_r \frac{R_{j,0,0}^{2r-1}}{2^{2r-1}} \left[ \sum_{l=0}^{2r-1} \binom{2r-1}{l} \int_0^t e^{i(\omega_i t - (2r-1-2l)(\omega_j t))} \right]$$

If i ≠ j, we will get the following,

$$\int_0^t W_{i,1,0} = \omega_i \sum g_{ij} \sum_{r=1}^{\infty} A_r \frac{R_{j,0,0}^{2r-1}}{2^{2r-1}} \left[ \sum_{l=0}^{2r-1} \binom{2r-1}{l} \frac{(e^{i(\omega_i-(2r-1-2l)\omega_j)t} - 1)}{(\omega_i - (2r-1-2l)\omega_j)} \right]$$

Eq. 4 and 5 will then give the following,

$$R_{i,1,0}(t) = \int_0^t \frac{W_{i,1,0} + W_{i,1,0}^*}{2}$$
$$= \frac{\omega_i}{2} \sum g_{ij} \sum_{r=1}^{\infty} A_r \frac{R_{j,0,0}^{2r-1}}{2^{2r-1}}$$
$$* \left[ \sum_{l=0}^{2r-1} \binom{2r-1}{l} \frac{\left(e^{i(\omega_i-(2r-1-2l)\omega_j)t} + e^{-i(\omega_i-(2r-1-2l)\omega_j)t}\right) - 2}{(\omega_i - (2r-1-2l)\omega_j)} \right]$$

$$\varphi_{i,1,0}(t) = i\int_0^t \frac{W_{i,1,0} - W_{i,1,0}^*}{2R_{i,0,0}}$$

$$= \frac{i\omega_i}{2R_{i,0,0}} \sum g_{ij} \sum_{r=1}^{\infty} A_r \frac{R_{j,0,0}^{2r-1}}{2^{2r-1}} *$$

$$* \left[ \sum_{l=0}^{2r-1} \binom{2r-1}{l} \frac{(e^{i(\omega_i-(2r-1-2l)\omega_j)t} - e^{-i(\omega_i-(2r-1-2l)\omega_j)t})}{(\omega_i - (2r-1-2l)\omega_j)} \right]$$

We can them see that the following holds,

$$R_{i,1,0}(T) = \varphi_{i,1,0}(T) = 0$$

If i=j, we will get the following,

$$\int_0^t W_{i,1,0} = -\omega_i g_{ii} \sum_{r=1}^{\infty} A_r \frac{R_{i,0,0}^{2r-1}}{2^{2r-1}} \sum_{l=0, l \neq r-1}^{2r-1} \binom{2r-1}{l} \frac{e^{-i((2r-2)-2l)\omega_i t} - 1}{((2r-2)-2l)\omega_i t}$$

$$+ i\omega_i g_{ii} \sum_{r=1}^{\infty} A_r \frac{R_{i,0,0}^{2r-1}}{2^{2r-1}} \binom{2r-1}{r-1} t$$

We thus have, where 1$^{st}$ and 2$^{nd}$ indicate the terms on the RHS in the preceding eq.,

$$2R_{i,1,0} = 1st + conjugate$$

$$-2iR_{i,0,0}\varphi_{i,1,0} = (1st + 2nd) - conjugate$$

We can now see that the following holds,

$$\frac{\varphi_{i,1,0}(T)}{T} = -\frac{\omega_i}{2R_{i,0,0}} g_{ii} \sum_{r=1}^{\infty} A_r \frac{R_{i,0,0}^{2r-1}}{2^{2r-1}} \binom{2r-1}{r-1}$$

$$R_{i,1,0}(T) = 0$$

In summary the topology of the phase space dynamics is stable for perturbations of first order in $\varepsilon$. The effect of the perturbation will be a change in the speed of rotation $\omega_i$.

$$\overline{\omega}_i = \omega_i \left(1 - \frac{1}{2R_{i,0,0}} g_{ii} \sum_{r=1}^{\infty} A_r \frac{R_{j,0,0}^{2r-1}}{2^{2r-1}} \binom{2r-1}{r-1}\right)$$

*A1.2 Expanding the $\varepsilon^2$ term*

Keeping only $\varepsilon^2$ terms for the LHS of Eq. A1 will give,

$$LHS = \left[\left(\dot{R}_{i,2,0} - iR_{i,0,0}\dot{\varphi}_{i,2,0}\right) + \frac{d}{dt}\left[-iR_{i,1,0}\varphi_{i,1,0} - \frac{R_{i,0,0}\varphi_{i,1,0}^2}{2}\right]\right] e^{-i\omega_i t}$$

Integrating this term after multiplying with $e^{i\omega_i t}$ will give,

$$\left(R_{i,2,0}(t) - iR_{i,0,0}\varphi_{i,2,0}(t)\right) - iR_{i,1,0}\varphi_{i,1,0} - \frac{R_{i,0,0}\varphi_{i,1,0}^2}{2}$$

The RHS of Eq. A1 (Eq A3) after multiplying with $e^{i\omega_i t}$ will give,

$$= i\varepsilon^1 \omega_i \sum g_{ij} \sum_{r=1}^{\infty} A_r \frac{\left(R_{j,0,0} + \Sigma \varepsilon^n \mu^m R_{j,n,m}(t)\right)^{2r-1}}{2^{2r-1}}$$

$$* \left[\sum_{l=0}^{2r-1} \binom{2r-1}{l} e^{-i((2r-1)-2l)(\omega_j t)+i\omega_i t} \left[\sum_{k=0}^{\infty} \frac{(-i(2r-1-2l)\varphi_j)^k}{k!}\right]\right]$$

Keeping only $\varepsilon^2$ terms will give,

$$= i\omega_i \sum g_{ij} \sum_{r=1}^{\infty} A_r \frac{R_{j,0,0}^{2r-1}}{2^{2r-1}} \quad \text{Eq. A6}$$

$$* \left[\sum_{l=0}^{2r-1} \binom{2r-1}{l} e^{-i((2r-1)-2l)(\omega_j t)+i\omega_i t} \left(-i(2r-1-2l)\varphi_{j,1,0}\right)\right]$$

$$+ i\omega_i \sum g_{ij} \sum_{r=1}^{\infty} A_r \frac{(2r-1)R_{j,0,0}^{2r-2} R_{j,1,0}}{2^{2r-1}}$$

$$* \left[\sum_{l=0}^{2r-1} \binom{2r-1}{l} e^{-i((2r-1)-2l)(\omega_j t)+i\omega_i t}\right]$$

Note that,

$$\varphi_{j,1,0} = \frac{i\omega_j}{2R_{i,0,0}} \sum_{j \neq k} g_{jk} \ldots + g_{jj}t \ldots$$

$$R_{j,1,0} = \frac{\omega_j}{2} \sum g_{jk} \ldots$$

Both functions will have a sum of products of exponentials. If none of the indices; i, j or k are equal, integrating this over a period T will not result in a net change. I.e. the cross terms will not contribute to a change in location of the trajectories over one full period T. The other four possible situations will be,

$$i = j \neq k$$
$$i = k \neq j$$
$$j = k \neq i$$
$$i = j = k$$

*If $i = j \neq k$*
The first term in Eq. A6 will contain the following term,

$$e^{\pm i(\omega_i - (2r-1-2l)\omega_k)t}$$

This term will not contribute to a change when integrated over a full cycle (t=T). The second term will contain the following terms,

$$e^{\pm i(\omega_i - (2r-1-2l)\omega_k)t} \text{ and } 1$$

The constant term will contribute to a change in the trajectory when integrated over a full cycle but will not change *R* but only $\varphi$ as the change will be imaginary. A similar situation will apply when $i = k \neq j$.

*If $j = k \neq i$*
The first term in Eq A6 will contain the following term,

$$\ldots e^{\pm i(\omega_i - (2r-1-2l)\omega_j)t} g_{jj}t \ldots$$

When this term is integrated using partial integration over the full cycle there will be an imaginary component at T, which will not contribute to a change in *R*. The terms from the second part will all have the following term,

$$e^{-i\omega_i t}$$

And will hence not have a non-zero contribution over a full cycle (t=T).

*If* $j = k = i$

The 2$^{nd}$ term in Eq. A6 will have an imaginary contribution and hence not change *R*. The 1$^{st}$ term will be given by,

$$= i\omega_i g_{ii} \sum_{r=1}^{\infty} A_r \frac{R_{i,0,0}^{2r-1}}{2^{2r-1}} * \left[ \sum_{l=0}^{2r-1} \binom{2r-1}{l} e^{-i((2r-2)-2l)(\omega_i t)} \left(-i(2r-1-2l)\varphi_{i,1,0}\right) \right]$$

Where,

Eq. A7

$$\varphi_{i,1,0}(t)$$
$$= -i\frac{\omega_i g_{ii}}{2R_{i,0,0}} \sum_{s=1}^{\infty} A_s \frac{R_{i,0,0}^{2s-1}}{2 * 2^{2s-1}} \sum_{m=0, m \neq s-1}^{2s-1} \binom{2s-1}{m} \frac{e^{-i((2s-2)-2m)\omega_i t} - e^{i((2s-2)-2m)\omega_i t}}{((2s-2)-2m)\omega_i t}$$
$$- \frac{\omega_i g_{ii}}{R_{i,0,0}} \sum_{s=1}^{\infty} A_s \frac{R_{i,0,0}^{2s-1}}{2^{2s-1}} \binom{2s-1}{s-1} t$$

The last term in Eq. 7 will give a real contribution at T if,

$$(2r-2) - 2l = 0$$

We then have for the 1st term in Eq. 6,

$$\omega_i g_{ii} \sum_{r=1}^{\infty} A_r \frac{R_{i,0,0}^{2r-1}}{2^{2r-1}} \binom{2r-1}{r-1} \varphi_{i,1,0}$$

Integrating this term will give,

$$-\omega_i g_{ii} \sum_{r=1}^{\infty} A_r \frac{R_{i,0,0}^{2r-1}}{2^{2r-1}} \binom{2r-1}{r-1} \frac{\omega_i g_{ii}}{R_{i,0,0}} \sum_{s=1}^{\infty} A_s \frac{R_{i,0,0}^{2s-1}}{2 * 2^{2s-1}} \binom{2s-1}{s-1} T^2 = -\frac{R_{i,0,0} \varphi_{i,1,0}(T)^2}{2}$$

This term will cancel the non-zero term on the LHS of Eq. A2.

$$-\frac{R_{i,0,0} \varphi_{i,1,0}^2}{2}$$

The 1st term in Eq. 7 will have an imaginary contribution and hence not change R. We have thus shown that the topology of the phase-space trajectories are stable for perturbations upto 2$^{nd}$ order.

## A2. Spike rate coupling
We will now investigate the effect of the spike rate coupling, i.e we will keep $\varepsilon = 0$ and analyse the effect of $\mu$ on the dynamics.

*A2.1 Expanding $\mu^1$ term,*

Keeping only $\mu^1$ terms for the LHS of Eq. A1 (Eq. A2) will give,

$$\mu^1 \dot{R}_{i,0,1}(t) e^{-i\omega_i t} - i R_{i,0,0} \mu^1 \dot{\varphi}_{i,0,1}(t) e^{-i\omega_i t}$$

Keeping only $\mu^1$ terms for the RHS of Eq. A1 (Eq. A3) will give,

$$-\mu \omega_i^2 \sum g_{ij} \sum_{r=1}^{\infty} (-1)^r B_r \frac{R_{j,0,0}^{2r-1}}{2^{2r-1}} * \left[ \sum_{l=0}^{2r-1} (-1)^l \binom{2r-1}{l} e^{-i((2r-1)-2l)(\omega_j t)} \right]$$

Equating these terms will give,

$$\dot{R}_{i,0,1}(t) e^{-i\omega_i t} - i R_{i,0,0} \dot{\varphi}_{i,0,1}(t) e^{-i\omega_i t}$$
$$= -\omega_i \sum g_{ij} \sum_{r=1}^{\infty} (-1)^r B_r \frac{R_{j,0,0}^{2r-1}}{2^{2r-1}}$$
$$* \left[ \sum_{l=0}^{2r-1} (-1)^l \binom{2r-1}{l} e^{-i((2r-1)-2l)(\omega_j t)} \right]$$

$$\dot{R}_{i,0,1}(t) - i R_{i,0,0} \dot{\varphi}_{i,0,1}(t)$$
$$= -\omega_i \sum g_{ij} \sum_{r=1}^{\infty} (-1)^r B_r \frac{R_{j,0,0}^{2r-1}}{2^{2r-1}}$$
$$* \left[ \sum_{l=0}^{2r-1} (-1)^l \binom{2r-1}{l} e^{-i((2r-1)-2l)(\omega_j t)+i\omega_i t} \right] = W_{i,0,1}$$

$$\dot{R}_{i,0,1}(t) = \frac{W_{i,0,1} + W_{i,0,1}^*}{2}$$

$$\dot{\varphi}_{i,0,1}(t) = i \frac{W_{i,0,1} - W_{i,0,1}^*}{2 R_{i,0,0}}$$

Integrating the terms will give,

$$R_{i,0,1}(t) = \int_0^t \dot{R}_{i,0,1}(t') = \int_0^t \frac{W_{i,0,1} + W_{i,0,1}^*}{2}$$

$$\varphi_{i,0,1}(t) = \int_0^t \dot{\varphi}_{i,0,1}(t') = i\int_0^t \frac{W_{i,0,1} - W_{i,0,1}{}^*}{2R_{i,0,0}}$$

The integral on *W* will be given by,

$$\int_0^t W_{i,1,0} = -\omega_i \sum g_{ij} \sum_{r=1}^{\infty}(-1)^r B_r \frac{R_{j,0,0}{}^{2r-1}}{2^{2r-1}}$$
$$* \left[\sum_{l=0}^{2r-1}(-1)^l \binom{2r-1}{l} \int e^{-i((2r-1)-2l)(\omega_j t)+i\omega_i t}\right]$$

If i ≠ j, we will get the following,

$$\int_0^t W_{i,0,1} = i\omega_i \sum g_{ij} \sum_{r=1}^{\infty}(-1)^r B_r \frac{R_{j,0,0}{}^{2r-1}}{2^{2r-1}} \left[\sum_{l=0}^{2r-1}(-1)^l \binom{2r-1}{l} \frac{(e^{i(\omega_i - (2r-1-2l)\omega_j)t} - 1)}{(\omega_i - (2r-1-2l)\omega_j)}\right]$$

$$R_{i,0,1}(t) = \int_0^t \frac{W_{i,0,1} + W_{i,0,1}{}^*}{2}$$
$$= i\omega_i \sum g_{ij} \sum_{r=1}^{\infty}(-1)^r B_r \frac{R_{j,0,0}{}^{2r-1}}{2^{2r}} \left[\sum_{l=0}^{2r-1}(-1)^l \binom{2r-1}{l} \frac{(e^{i(\omega_i - (2r-1-2l)\omega_j)t} - e^{-i(\omega_i - (2r-1-2l)\omega_j)t})}{(\omega_i - (2r-1-2l)\omega_j)}\right]$$

$$\varphi_{i,0,1}(t) = i\int_0^t \frac{W_{i,0,1} - W_{i,0,1}{}^*}{2R_{i,0,0}}$$
$$= i\omega_i \sum g_{ij} \sum_{r=1}^{\infty}(-1)^r B_r \frac{R_{j,0,0}{}^{2r-1}}{2^{2r} R_{i,0,0}} \left[\sum_{l=0}^{2r-1}(-1)^l \binom{2r-1}{l} \frac{(ie^{i(\omega_i - (2r-1-2l)\omega_j)t} + ie^{-i(\omega_i - (2r-1-2l)\omega_j)t} - 2i)}{(\omega_i - (2r-1-2l)\omega_j)}\right]$$

We have shown that the off diagonal terms (i ≠ j) do not change the topology of the phase space structure upto first order, i.e.,

$$R_{i,0,1}(T) = \varphi_{i,0,1}(T) = 0$$

If i=j we will have,

$$\int_0^t W_{i,0,1} = -\omega_i g_{ii} \sum_{r=1}^{\infty}(-1)^r B_r \frac{R_{i,0,0}{}^{2r-1}}{2^{2r-1}} * \left[\sum_{l=0}^{2r-1}(-1)^l \binom{2r-1}{l} \int e^{-i((2r-2)-2l)(\omega_i t)}\right]$$

$$= -t\omega_i g_{ii} \sum_{r=1}^{\infty}(-1)^r(-1)^{r-1}B_r \frac{R_{i,0,0}^{2r-1}}{2^{2r-1}}\binom{2r-1}{r-1}$$

$$- ig_{ii}\sum_{r=1}^{\infty}(-1)^r B_r \frac{R_{i,0,0}^{2r-1}}{2^{2r-1}}$$

$$* \left[\sum_{l\neq r-1, l=0}^{2r-1}(-1)^l\binom{2r-1}{l}\frac{e^{-i((2r-2)-2l)(\omega_i t)}-1}{((2r-2)-2l)}\right]$$

$$R_{i,1}(T) = \mu T \omega_i g_{ii} \sum_{r=1}^{\infty} B_r \frac{R_{i,0,0}^{2r-1}}{2^{2r-1}}\binom{2r-1}{r-1}$$

Hence self-interaction terms create changes in the R coordinate as the flow completes full cycles. The rate of change in *R* can be estimated as,

$$\frac{dR_i}{dt} = \frac{\delta R_i}{T} = \frac{\mu R_{i,0,1}(T)}{T} = \mu \omega_i g_{ii} \sum_{r=1}^{\infty} B_r \frac{R_{i,0,0}^{2r-1}}{2^{2r-1}}\binom{2r-1}{r-1}$$

The zeroes of the above function will correspond to stationary points or limits cycles. E.g.

$$-\mu\omega_i g_{ii}\left(4R_{i,0,0} - 5R_{i,0,0}^3 + R_{i,0,0}^5\right)$$

The above equation will have stationary points or limit cycles at 0,1 and 2. The corresponding value of $B_r$ will be given by,

$$B_r = \{8, -\frac{40}{3}, \frac{16}{5}\}$$

N-stable limit cycles and a stationary point at the origin could be given using the following equation,

$$-\mu\omega_i g_{ii} R_{i,0,0}(R_{i,0,0}^2 - a_1^2)\dots(R_{i,0,0}^2 - a_{2N}^2)$$

$$= -\mu\omega_i g_{ii}\left[(-1)^{2N}R_{i,0,0}\prod_{i=1}^{2N}a_i^2 + (-1)^{2N-1}R_{i,0,0}^3 \sum_{j=1}^{2N}\frac{1}{a_j^2}\prod_{i=1}^{2N}a_i^2 + \cdots\right.$$

$$+ (-1)^{2N-n}R_{i,0,0}^1 R_{i,0,0}^{2n} \sum_{1\leq j_1<\cdots<a_{j_n}\leq 2N}\frac{1}{a_{j_1}^2\dots a_{j_n}^2}\prod_{i=1}^{2N}a_i^2 + \cdots$$

$$\left. + R_{i,0,0}^{4N+1}\right]$$

The corresponding values of $B_r$ will be given by,

$$B_{n+1} = \frac{(-1)^{2N-n} 2^{2n+1} R_{i,0}^{2n+1}}{\binom{2n+1}{n}} \prod_{i=1}^{2N} a_i^2 \sum_{1 \leq j_1 < \cdots < j_n \leq 2N}^{P(2N,n)} \frac{1}{a_{j_1}^2 \ldots a_{j_n}^2}$$

We have used $P(2N, n)$ to indicate that the sum is taken over all combinations of n index points from a set of 2N index points. The dynamics of *R* can be written as a potential flow.

$$\frac{dR_i}{dt} = -\frac{dU}{dR_i}$$

$$U = -\mu \omega_i g_{ii} \sum_{r=1}^{\infty} B_r \frac{R_{i,0,0}^{2r}}{r 2^{2r}} \binom{2r-1}{r-1}$$

*A2.2 Expanding the $\mu^2$ term*

Keeping only $\mu^2$ terms gives for the LHS of Eq. A1 (Eq. A2),

$$LHS = \left[ \left( \dot{R}_{i,0,2} - i R_{i,0,0} \dot{\varphi}_{i,0,2} \right) + \frac{d}{dt}\left[ -i R_{i,0,1} \varphi_{i,0,1} - \frac{R_{i,0,0} \varphi_{i,0,1}^2}{2} \right] \right] e^{-i\omega_i t}$$

Integrating this term after multiplying with $e^{i\omega_i t}$ will give,

$$\left[ \left( R_{i,2}(t) - i R_{i,0,0} \varphi_{i,0,2}(t) \right) - i R_{i,0,1} \varphi_{i,0,1} - \frac{R_{i,0,0} \varphi_{i,0,1}^2}{2} \right]\Big|_0^t$$

The RHS of Eq A3 will give after multiplying with $e^{i\omega_i t}$,

$$= -\mu^1 \omega_i \sum g_{ij} \sum_{r=1}^{\infty} (-1)^r B_r \frac{\left( R_{j,0,0} + \sum \mu^n R_{j,0,n}(t) \right)^{2r-1}}{2^{2r-1}}$$

$$* \left[ \sum_{l=0}^{2r-1} (-1)^l \binom{2r-1}{l} e^{-i((2r-1)-2l)(\omega_j t)+i\omega_i t} \left[ \sum_{k=0}^{\infty} \frac{\left(-i(2r-1-2l)\varphi_j\right)^k}{k!} \right] \right]$$

Keeping only $\mu^2$ terms will give,

$$= -\omega_i \sum g_{ij} \sum_{r=1}^{\infty} (-1)^r B_r \frac{R_{j,0,0}^{2r-1}}{2^{2r-1}} \quad \text{Eq. A8}$$

$$* \left[ \sum_{l=0}^{2r-1} (-1)^l \binom{2r-1}{l} e^{-i((2r-1)-2l)(\omega_j t)+i\omega_i t} * \left(-i(2r-1-2l)\varphi_{j,0,1}\right)\right]$$

$$- \omega_i \sum g_{ij} \sum_{r=1}^{\infty} (-1)^r B_r \frac{(2r-1)R_{j,0,0}^{2r-2} R_{j,0,1}}{2^{2r-1}}$$

$$* \left[ \sum_{l=0}^{2r-1} (-1)^l \binom{2r-1}{l} e^{-i((2r-1)-2l)(\omega_j t)+i\omega_i t} \right]$$

The first order components are given by (if $i \neq j$),

$$R_{i,0,1}(t) = i\omega_i \sum g_{ij} \sum_{r=1}^{\infty} (-1)^r B_r \frac{R_{j,0,0}^{2r-1}}{2^{2r}} \left[ \sum_{l=0}^{2r-1} (-1)^l \binom{2r-1}{l} \frac{(e^{i(\omega_i-(2r-1-2l)\omega_j)t} - e^{-i(\omega_i-(2r-1-2l)\omega_j)t})}{(\omega_i - (2r-1-2l)\omega_j)} \right]$$

$$\varphi_{i,0,1}(t) = \omega_i \sum g_{ij} \sum_{r=1}^{\infty} (-1)^r B_r \frac{R_{j,0,0}^{2r-1}}{2^{2r} R_{i,0,0}} \left[ \sum_{l=0}^{2r-1} (-1)^l \binom{2r-1}{l} \frac{(e^{i(\omega_i-(2r-1-2l)\omega_j)t} + e^{-i(\omega_i-(2r-1-2l)\omega_j)t} - 2)}{(\omega_i - (2r-1-2l)\omega_j)} \right]$$

The diagonal component (if $i = j$ will be given by,

$$R_{i,0,1}(t) = -t\omega_i g_{ii} \sum_{r=1}^{\infty} (-1)^r (-1)^{r-1} B_r \frac{R_{i,0,0}^{2r-1}}{2^{2r-1}} \binom{2r-1}{r-1}$$

$$- ig_{ii} \sum_{r=1}^{\infty} (-1)^r B_r \frac{R_{i,0,0}^{2r-1}}{2^{2r-1}}$$

$$* \left[ \sum_{l \neq r-1, l=0}^{2r-1} (-1)^l \binom{2r-1}{l} \frac{e^{-i((2r-2)-2l)(\omega_i t)} - e^{i((2r-2)-2l)(\omega_i t)}}{2((2r-2)-2l)} \right]$$

$$\varphi_{i,0,1}(t) = g_{ii} \sum_{r=1}^{\infty} (-1)^r B_r \frac{R_{i,0,0}^{2r-2}}{2^{2r-1}}$$

$$* \left[ \sum_{l \neq r-1, l=0}^{2r-1} (-1)^l \binom{2r-1}{l} \frac{e^{-i((2r-2)-2l)(\omega_i t)} + e^{i((2r-2)-2l)(\omega_i t)} - 2}{2((2r-2)-2l)} \right]$$

The following conditions will need to be used to expand the equations for the 2nd order contribution to $R$ and $\varphi$ (the indices will run over 3 subpopulations of neurons, i, j and k):

1) $i = j = k$

We will then have the following for the RHS,

$$= -\omega_i g_{ii} \sum_{r=1}^{\infty} (-1)^r B_r \frac{R_{i,0,0}^{2r-1}}{2^{2r-1}}$$

$$* \left[ \sum_{l=0}^{2r-1} (-1)^l \binom{2r-1}{l} e^{-i((2r-2)-2l)(\omega_i t)} * \left(-i(2r-1-2l)\varphi_{i,0,1}\right) \right]$$

$$- \omega_i g_{ii} \sum_{r=1}^{\infty} (-1)^r B_r \frac{(2r-1)R_{i,0,0}^{2r-2} R_{i,0,1}}{2^{2r-1}}$$

$$* \left[ \sum_{l=0}^{2r-1} (-1)^l \binom{2r-1}{l} e^{-i((2r-2)-2l)(\omega_i t)} \right]$$

Eq. A9

The first term in Eq. A9 will give the following,

$$-\omega_i g_{ii} \sum_{r=1}^{\infty} (-1)^r B_r \frac{R_{i,0,0}^{2r-1}}{2^{2r-1}}$$

$$* \left[ \sum_{l=0}^{2r-1} (-1)^l \binom{2r-1}{l} e^{-i((2r-2)-2l)(\omega_i t)} * \left(-i(2r-1-2l)\varphi_{i,0,1}\right) \right]$$

There will only be imaginary contributions which will not affect $R$. The second term in Eq. A9 will give the following,

$$-\omega_i g_{ii} \sum_{r=1}^{\infty} (-1)^r B_r \frac{(2r-1)R_{i,0,0}^{2r-2} R_{i,0,1}}{2^{2r-1}} * \left[ \sum_{l=0}^{2r-1} (-1)^l \binom{2r-1}{l} e^{-i((2r-2)-2l)(\omega_i t)} \right]$$

The only contribution to $R$ will be given by,

$$\omega_i^2 g_{ii}^2 T \sum_{r=1}^{\infty} (-1)^r B_r \frac{(2r-1)R_{i,0,0}^{2r-2}}{2^{2r-1}} * \left[ \sum_{l=0}^{2r-1} (-1)^{r-1} \binom{2r-1}{r-1} \right]$$

$$* \sum_{s=1}^{\infty} (-1)^s (-1)^{s-1} B_s \frac{R_{i,0,0}^{2s-1}}{2^{2s-1}} \binom{2s-1}{s-1}$$

$$= \omega_i^2 g_{ii}^2 T \sum_{r,s=1}^{\infty} B_r B_s \frac{(2r-1)R_{i,0,0}^{2r+2s-3}}{2^{2r-1} 2^{2s-1}} \binom{2r-1}{r-1}\binom{2s-1}{s-1}$$

2) $i = j \neq k$

When $i = j \neq k$ we will only get imaginary components which will not contribute to *R*.

3) $i \neq j = k$

When $i \neq j = k$ we will get only get imaginary components which will not contribute to *R*.

4) $i \neq j, j \neq k, i \neq k$

This condition ($i \neq j, j \neq k, i \neq k$ ) will be not contribute to *R* over a full cycle (*T*).

5) $i \neq j, i = k$

The RHS (Eq. A8) will give,

$$= -\omega_i \sum g_{ij} \sum_{r=1}^{\infty} (-1)^r B_r \frac{R_{j,0,0}^{2r-1}}{2^{2r-1}}$$
$$* \left[ \sum_{l=0}^{2r-1} (-1)^l \binom{2r-1}{l} e^{-i((2r-1)-2l)(\omega_j t)+i\omega_i t} * \left(-i(2r-1-2l)\varphi_{j,1}\right) \right]$$
$$- \omega_i \sum g_{ij} \sum_{r=1}^{\infty} (-1)^r B_r \frac{(2r-1)R_{j,0,0}^{2r-2} R_{j,0,1}}{2^{2r-1}}$$
$$* \left[ \sum_{l=0}^{2r-1} (-1)^l \binom{2r-1}{l} e^{-i((2r-1)-2l)(\omega_j t)+i\omega_i t} \right]$$

Again the terms contributing over a whole cycle will be imaginary and will not affect *R*.

In summary we have shown that potential-to-current coupling will not change the topology of the phase space structure, while the direct self-interaction terms of the current-to-current will.

A3. Cross coupling between potential-to-current and current-to-current coupling

It can be shown using a similar procedure to the above derivations that cross coupling between potential-to-current and current-to-current coupling will cause an instability in the phase space structure. In contrast, to what was shown in A1-2, off-diagonal terms ($g_{ij}, i \neq j$) will contribute to changes in R.

$$\frac{dR_{i,1,1}}{dt} = f(g_{ij}, R_{i,0,0}, R_{j,0,0}, S, P)$$

## A4. Amplitude envelope dynamics

The lead field of the EEG (Eq. 9) defines ellipsoidal surfaces in phase space where the amplitude power of the EEG will be constant, Figure A1. As described in the main text we will perform a spatial average in phase space over these surfaces. The resultant spatially averaged amplitude will be a function of the distance to the origin.

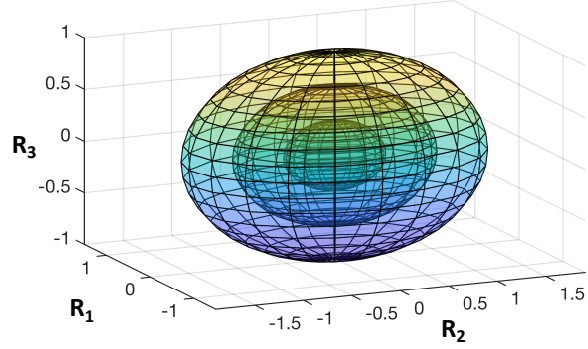

**Figure A1. 2D-ellipoids representing the surfaces with constant EEG amplitude.** The spatial average of the amplitude flow will be done over the positive sector ($R_i>0$) of each of these surfaces.

Hyperspheroidal coordinates (in $\mathbb{R}^4$) can be used for the integration,

$$R_1 = \frac{R}{a_1} cos\phi_1$$

$$R_2 = \frac{R}{a_2} sin\phi_1 cos\phi_2$$

$$R_3 = \frac{R}{a_3} sin\phi_1 sin\phi_2 cos\phi_3$$

$$R_4 = \frac{R}{a_4} sin\phi_1 sin\phi_2 sin\phi_3$$

The area unit for the ellipsoid is given by,

$$dE \equiv dE(\boldsymbol{a}, \boldsymbol{\phi})$$

This is a complicated function of the coordinates ($\boldsymbol{\phi}$) and the components of the linear field ($\boldsymbol{a}$). Again, we will simplify to be able to integrate analytically. Instead of the ellipsoidal area unit we use the spheroidal area unit ($dA$), see figure A2. For a 3D-sphere this simplifies to,

$$dA = R^3 sin^2\phi_1 sin\phi_2 d\phi_1 d\phi_2 d\phi_3$$

$$A = \frac{1}{4}\pi^2 R^3$$

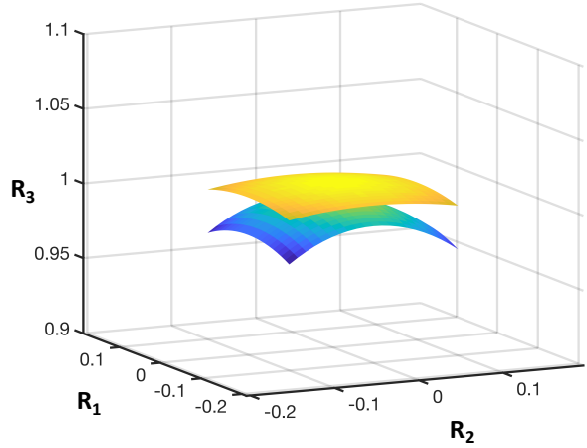

**Figure A2. 2D-spherical and 2D-ellipsoidal area unit.** The spherical (yellow) and the ellipsoidal (blue) area unit are separated for better visualization. Using the spherical area unit instead of the elliptic will simplify the integrals considerably allowing for analytical solutions.

Note the following integral over $\frac{\pi}{2}$:

$$\int \sin^m \phi_1 \cos^n \phi_1 d\phi_1 = PI = \frac{\sin^{m+1}\phi_1}{m+1}\cos^{n-1}\phi_1 + \frac{n-1}{m+1}\int \sin^{m+2}\phi_1 \cos^{n-2}\phi_1$$

$$\int \sin^m \phi_1 \cos^n \phi_1 d\phi_1 = \frac{(n-1)!!\,(m-1)!!}{(m+n-1)!!}\int \sin^{m+n}\phi_1$$

The following terms need to be evaluated,

$$\int R_i \frac{dR_i}{dt} dA = \mu \omega_i g_{ii} \int \sum_{r=1}^{\infty} A_r \frac{R_{i,0,0}^{2r}}{2^{2r-1}} \binom{2r-1}{r-1} dA$$

We use the spherical area unit as an approximation to the ellipsoidal area unit to evaluate the requisite integrals. The self-interaction of the 1st population gives the following,

$$\mu \omega_1 g_{11} \binom{2r-1}{r-1} \sum_{r=1}^{\infty} \frac{B_r}{2^{2r-1}} \int R_{1,0,0}^{2r} dA = \cdots \int R_{1,0}^{2r} dA$$

$$\int R_{1,0,0}^{2r} dA = \frac{R^{2r+3}}{a_1^{2r}} \int \cos^{2r}\phi_1 (1-\cos^2\phi_1)\sin\phi_2 d\phi_1 d\phi_2 d\phi_3$$

$$= \frac{R^{2r+3}}{a_1^{2r}} \int \cos^{2r}\phi_1 - \cos^{2r+2}\phi_1 d\phi_1 \int \sin\phi_2 d\phi_2 \int d\phi_3$$

Integral over $d\phi_1$,

$$\int \cos^{2r}\phi_1 - \cos^{2r+2}\phi_1 d\phi_1 = \frac{\pi}{2(2r+1)} \frac{(2r+1)!!}{(2r+2)!!}$$

Integral over $d\phi_2$,

$$\int \sin\phi_2 d\phi_2 = 1$$

Integral over $d\phi_2$,

$$\int d\phi_3 = \frac{\pi}{2}$$

Full integral

$$\int R_{1,0}{}^{2r} dA = \frac{4\pi^2 R^{2r+3}}{(2r+1)a_1{}^{2r}} \frac{(2r+1)!!}{(2r+2)!!}$$

From symmetry we get similar terms for the other populations.

$$\int R_i \frac{dR_i}{dt} dA = \mu\omega_i g_{ii} \sum_{r=1}^{\infty} B_r \frac{4\pi^2 R^{2r+3}}{(2r+1)2^{2r}a_i{}^{2r}} \binom{2r-1}{r-1}\frac{(2r+1)!!}{(2r+2)!!}$$

Let $R_s$ be the average of $R$ over the positive sector of the 3-sphere,

$$\frac{dI_s}{dt} = \frac{1}{AR} \oiint \sum a_i{}^2 R_i \frac{dR_i}{dt} = \mu \sum a_i{}^2 \omega_i g_{ii} \sum_{r=1}^{\infty} A_r \frac{4R^{2r-1}}{(2r+1)2^{2r}a_i{}^{2r}} \binom{2r-1}{r-1}\frac{(2r+1)!!}{(2r+2)!!} \qquad \text{Eq. A10}$$